\documentstyle[mncite,draft,epsf]{mn}

\newcommand{\vsini}{\mbox{$v_e\,\sin\,i$}}

\newcommand{\kms}{\,km\,s$^{-1}$}

\newcommand{\msun}{\,\mbox{$\mbox{M}_{\odot}$}}

\newcommand{\be}{\begin{equation}}
\newcommand{\ee}{\end{equation}}
\newcommand{\bd}{\begin{displaymath}}
\newcommand{\ed}{\end{displaymath}}

\title[Rotation and activity in NGC 2547]{
Rotation and activity in the solar-type stars of NGC 2547}  
  
\author[R. D. Jeffries et al.] {R. D. Jeffries$^{1}$,  
E. J. Totten$^{1}$, D. J. James$^{2,3}$ \\  
$^{1}$Department of Physics, Keele University, Keele, Staffordshire,  
ST5 5BG, UK\\  
$^{2}$ Department of Physics and Astronomy, University of St Andrews,  
Fife, KY16 9SS, UK\\  
$^{3}$ Observatoire de Gen\`{e}ve, Chemin de Maillettes 51, CH-1290,
Sauverny, Switzerland}

\date{Received 10 February 2000}  
\pagerange{\pageref{firstpage}--\pageref{lastpage}}  
\def\LaTeX{L\kern-.36em\raise.3ex\hbox{a}\kern-.15em  
    T\kern-.1667em\lower.7ex\hbox{E}\kern-.125emX}

\begin{document}  
  
\label{firstpage}  
  
\maketitle  
  
\begin{abstract}  
We present high resolution spectroscopy of a sample of 24 solar-type stars
in the young (15-40\,Myr), open cluster, NGC 2547. We use our spectra to confirm
cluster membership in 23 of these stars, determine
projected equatorial velocities and chromospheric activity, and to search
for the presence of accretion discs. We have found examples of both
fast (\vsini$>50$\kms) and slow (\vsini$<10$\kms) rotators, but find no
evidence for active accretion in any of the sample. The distribution of
projected rotation velocities is indistinguishable from the slightly
older IC 2391 and IC 2602 clusters, implying similar initial angular
momentum distributions and circumstellar disc lifetimes. The presence
of very slow rotators indicates that either long (10-40\,Myr) disc
lifetimes or internal differential rotation are needed, or that NGC 2547 (and IC
2391/2602) were born with more slowly rotating stars than are presently seen
in even younger clusters and associations. The solar-type stars in NGC 2547
follow a similar rotation-activity relationship to that seen in older
clusters. X-ray activity increases until a saturation level is reached for
$\vsini>15-20$\kms. We are unable to explain why this saturation level,
of $\log (L_{\rm x}/L_{\rm bol})\simeq -3.3$, is a factor of
two lower than in other clusters, but rule out anomalously slow
rotation rates or uncertainties in X-ray flux calculations.
\end{abstract}  
  
  
\section{Introduction}  
  
Open clusters are obvious laboratories in which to study the evolution
of stellar X-ray activity. They contain stars with a variety of masses
but a similar age, distance and composition. A large amount of {\em
Einstein}, and {\em ROSAT} X-ray observatory time was spent looking at
open clusters (see for instance the reviews of Randich 1997 and
Jeffries 1999) and a major achievement of these missions was to show
that solar analogues, and stars of even lower mass, in young open
clusters, could have X-ray activity orders of magnitude greater than
the Sun. This activity correlates with fast rotation and is
hypothesized to be due to an internal convective dynamo which generates
the magnetic fields that both heat and confine hot coronae. The
evolution of X-ray activity is thought to obey an age-rotation-activity
paradigm (ARAP). Young single stars are rapidly rotating and active,
but lose angular momentum and spin-down as they age, resulting in a
decrease of their X-ray activity.

The spin-down of cool stars as they age is not as simple as the $\Omega
\propto t^{-1/2}$ law once proposed by Skumanich (1972). It appears
that G and K stars stars arrive on the main sequence (for instance in
the Pleiades -- age $\simeq100$\,Myr) with a spread in rotation rates
from a few times to a hundred times the solar rotation
rate. Subsequently, this spread almost converges by the age (600\,Myr)
of the Hyades. In recent times these phenomena have been understood in
terms of braking caused by a magnetically coupled, ionized wind,
discussed in detail by (for instance) Barnes \& Sofia (1996) and
Krishnamurthi et al. (1997).  The spread in rotation rates observed at
the ZAMS combined with the magnitude and rather narrow distribution of
rotation rates among pre main-sequence (PMS) stars implies that, in
order to produce the fastest ZAMS rotators, substantial angular
momentum must be lost during the PMS phase, and the wind braking
mechanism must saturate at high rotation rates (Bouvier, Forestini \&
Allain 1997a).  This could be due to a saturation of the dynamo
mechanism itself, or perhaps changes in the magnetic field geometry
(Barnes \& Sofia 1996).

It has long been supposed that during the initial stages of PMS
rotation evolution, wind braking was negligible compared with the
moment of inertia decrease as a star contracts towards the ZAMS. An
alternative angular momentum loss mechanism is required and might be provided by
magnetic torques transferring angular momentum to a circumstellar disc
(K\"{o}nigl 1991; Shu et al. 1994). It can be shown that disc-regulated
angular momentum loss leads to almost constant angular velocity as a
PMS star shrinks.  A range of disc lifetimes, perhaps connected to the
initial disc mass (Armitage \& Clarke 1996), then results in stars
becoming decoupled from their discs at varying ages. Spin up from this
point to the ZAMS, combined with wind angular momentum loss that
saturates above some threshold rotation rate, can provide a very wide
spread of ZAMS rotation rates. This scenario has been modelled
extensively by Keppens, MacGregor \& Charbonneau (1995),
Collier-Cameron, Campbell \& Quaintrell (1995), Bouvier et
al. (1997a) and Krishnamurthi et al. (1997).

\begin{figure*}  
\vspace{11.0cm} \includegraphics{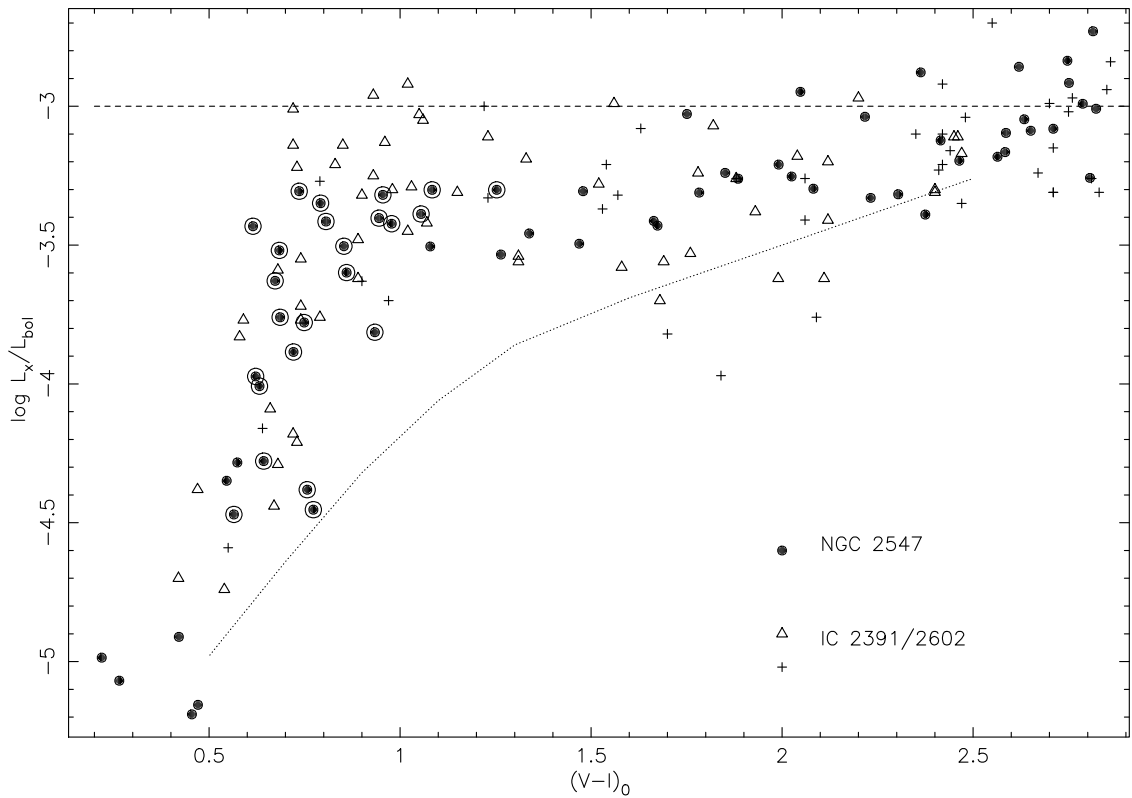}  
\caption{X-ray activity, expressed as $L_{\rm x}/L_{\rm bol}$, 
in NGC 2547 (dots) as a function of
intrinsic colour, compared with
stars in IC 2391/2602 (triangles and crosses). 
The dotted line indicates the
X-ray sensitivity limit for NGC 2547 and the dashed line indicates the
``saturation level'' for X-ray emission reached in all young clusters. 
The circled NGC 2547 points were selected for high resolution optical
spectroscopy in this paper. Triangles represent members of IC 2391/2602
confirmed by spectroscopy in S97. Crosses are X-ray selected members of
IC 2391/2602 so far unconfirmed by spectroscopy.}
\end{figure*}

Support for disc-regulated PMS angular momentum loss comes from
observations that show a connection between rotation rates in PMS stars
and the presence of circumstellar accretion discs. Choi \& Herbst
(1996 and references therein) claimed that rotation rates in the Orion
nebula cluster were bimodal, with periods either shorter or longer than
4-5 days. They proposed that the slower group of rotators might be
those that still maintained a circumstellar accretion disc. Bouvier et
al. (1993) showed that in their sample of Taurus-Auriga PMS stars, the
classical T-Tauri stars (CTTS - those stars showing optical signatures
of active accretion and circumstellar discs) rotated more slowly on
average than the weak T-Tauri stars (WTTS) that showed no sign of
circumstellar discs. Furthermore, Edwards et al. (1993) were able to
show that a similar situation held when the near-IR signatures of discs
were considered for Orion PMS stars. More recently the whole
disc-regulated angular momentum loss idea has been challenged by
Stassun et al. (1999), who find little correlation between disc
signatures and rotation rates in a larger sample of low mass
(0.2-0.6\msun) Orion PMS stars, as well as a population of extremely
rapid rotators (periods $<2$ day) both with and without discs. They
claim that an alternative to disc-locking is required, if the Orion
rotation rate distribution is to evolve into that seen among the ZAMS
stars of the Pleiades cluster.

These new ideas on the evolution of stellar rotation rate have been
successfully applied to X-ray observations of young clusters of stars
({\em e.g.} IC 2391 and IC 2602, age $\simeq 30$\,Myr -- Stauffer et
al. 1997 [S97], Alpha Per, age $\simeq 60$\,Myr -- Randich et al. 1996,
Pleiades, age $\simeq100$\,Myr -- Stauffer et al. 1994).  In every
cluster studied so far the ARAP appears to hold, with rotation
providing the parameter that determines the level of X-ray emission
from a star. This was most clearly demonstrated in the Pleiades by
Stauffer et al. (1994). They showed that X-ray activity (measured by
$L_{\rm x}/L_{\rm bol}$) in solar-type stars, increases with $\vsini$
until a saturation is reached at $L_{\rm x}/L_{\rm bol}\simeq 10^{-3}$
for $\vsini>15-20$\kms, with perhaps a decrease in the \vsini\
threshold for lower mass stars with thicker convection zones
(Krishnamurthi et al. 1998).  This relationship has since been
confirmed in the younger Alpha Per and IC 2391/2602 clusters, with some
hint that the X-ray activity may even decline at ultra-fast rotation
rates ($\vsini>100$\kms -- Randich et al. 1998).

NGC 2547 is an excellent cluster with which to explore some of these
issues. Jeffries \& Tolley (1998 [JT98]) reported X-ray observations of
the cluster, detecting a rich population of PMS stars with an age,
deduced from fits to low mass isochrones, of $(14\pm4)$\,Myr --
somewhat younger than IC 2391/2602. In order to produce the large
numbers ($\sim$85\% with $\vsini<20$\kms) of slow rotators at the age
of the Pleiades, the solid-body rotation models of Bouvier et
al. (1997a) predict that at $\sim15$\,Myr about 15-20\% of solar-type
stars should still possess circumstellar discs and that approximately
50\% of stars should have a projected equatorial velocity (\vsini) less
than 20\kms.  Shorter disc lifetimes are allowed if differential
rotation between the core and convective envelope is possible (Keppens
et al. 1995; Krishnamurthi et al. 1997), and would be more 
in accord with the (few) available
measurements in T-Tauri stars, which suggest median and maximum disc
lifetimes of about 1-3\,Myr and $\sim10$\,Myr respectively (Strom et
al. 1989; Skrutskie et al. 1990; Edwards et al. 1993; Kenyon \&
Hartmann 1995).

We would therefore expect (according to the ARAP) that approximately
half of the solar-type stars in NGC 2547 should show saturated X-ray emission.
In fact the distribution of X-ray emission in NGC 2547 does
not meet these expectations. Figure 1 compares $L_{\rm x}/L_{\rm bol}$
as a function of intrinsic colour for NGC 2547 with that of IC
2391/2602 members (taken from Randich et al. 1995; Patten \& Simon
1996; S97 and Simon \& Patten 1998). The X-ray activity appears to peak
or saturate in NGC 2547 at a level about a factor of two lower than in
IC 2391/2602 for the late F to early K stars. The same is true when NGC
2547 is compared with the Alpha Per and Pleiades clusters (see also
Fig.6 in Randich et al. 1995 and Fig. 13
in JT98).  JT98 argued that this was not a simple scaling error in the
conversion from X-ray count rates to X-ray fluxes in NGC 2547, because
cooler stars ($(B-V)_{0}>1.3$, $(V-I)_{0}>1.8$) appear to have a
saturation level, $L_{\rm x}/L_{\rm bol}\simeq 10^{-3}$, that is
consistent with other clusters. They suggested that perhaps {\em all} the
solar-type stars in NGC 2547 were rotating slower than a speed of
20\kms, above which, the X-ray emission of solar-type stars in other
clusters appears to saturate.  This might be the case if the stars in
NGC 2547 had either started life with much less angular momentum than
stars in other clusters or had somehow retained circumstellar accretion
discs for longer than usual.

The possibility that disc lifetimes could exceed 10\,Myr in some
circumstances would have important implications for stellar angular
momentum loss and the possible formation of planetary systems.  In this
paper we present the results of high resolution spectroscopy of a
sample of solar-type stars in NGC 2547. Our aim is to test for the
presence of active accretion discs, measure rotation rates and to see
whether the ARAP can successfully explain the X-ray emission that is
seen in these stars. In section 2 we outline the sample of stars we
have observed, the observations that were made and their
analysis. Section 3 presents the results of these analyses and compares
the rotation and activity in NGC 2547 with that in other young
clusters. These results are discussed in section 4.

\section{Observations}

\subsection{Sample selection}

All of our targets were selected as optical counterparts to X-ray
sources by JT98. They have $B-V$ and $V-I_{\rm c}$ 
colours and $V$ magnitudes that are consistent
with membership of the NGC 2547 cluster. For the remainder of this
paper we assume that the intrinsic distance modulus to the cluster is
$8.1\pm0.1$ and that the reddening is given by $E(B-V)=0.06$ or
$E(V-I_{\rm c})=0.077$ (see
JT98). An earlier photoelectric study of early-type members by
Clari\'{a} (1982) has established that any differential reddening in
the cluster is less than 0.02 (rms).
JT98 fitted isochrones to the X-ray sources in NGC 2547 and found an
age of $14\pm4$\,Myr, compared to about 25\,Myr for IC 2391 and IC 2602
using the same isochrones and colour-$T_{eff}$ relationship. There is
some evidence that these isochronal ages may be underestimates (see section 4.1).

Figure 1 shows the distribution of X-ray activity in NGC 2547 as a
function of intrinsic colour.
The approximate sensitivity threshold for detecting X-ray sources
in the cluster is shown as a dotted line. This threshold was determined
by JT98 and is appropriate for cluster members situated near the centre
of the {\em ROSAT} field of view. For objects nearer the edge of the
{\em ROSAT} field of view, the detection threshold is approximately 
0.3 dex higher. The argument used by JT98, which we shall also use
here, is that whilst an X-ray selected sample of cluster members would
normally be biased towards faster rotators; if the X-ray sensitivity
threshold is low enough that there are no cluster members situated at
or slightly above this threshold, then it is most likely that the X-ray
selected sample is complete. This might not be true if X-ray luminosity
functions showed a bimodal distribution, but
this is not the case in the solar-type stars of the
Pleiades and Hyades where complete optical samples have been observed
at X-ray wavelengths (Stern et al. 1992; Stauffer et al. 1994).

If we now consider just the stars in NGC 2547 with $0.8<(V-I_{\rm
c})_{0}<1.40$ -- corresponding roughly to masses $1.0\msun>M>0.8$\msun\ according
to the D'Antona \& Mazzitelli (1997) isochrones used to determine the
age, there does seem to be a significant gap between the X-ray
sensitivity threshold and the lowest activity levels detected,
indicating that the X-ray selected sample of solar-type stars should be
almost complete. The same may
not be true for hotter and cooler stars, where it is still possible
that some slow rotators have not been seen in X-rays. The number of
X-ray selected solar-type stars found in the cluster {\em is}
consistent with the number of higher mass stars and canonical initial
mass functions (see JT98 for details). However, the uncertainties in
these numbers are too large to place any strong constraints on the 
sample completeness.
\begin{figure}  
\vspace{10.0cm} \includegraphics{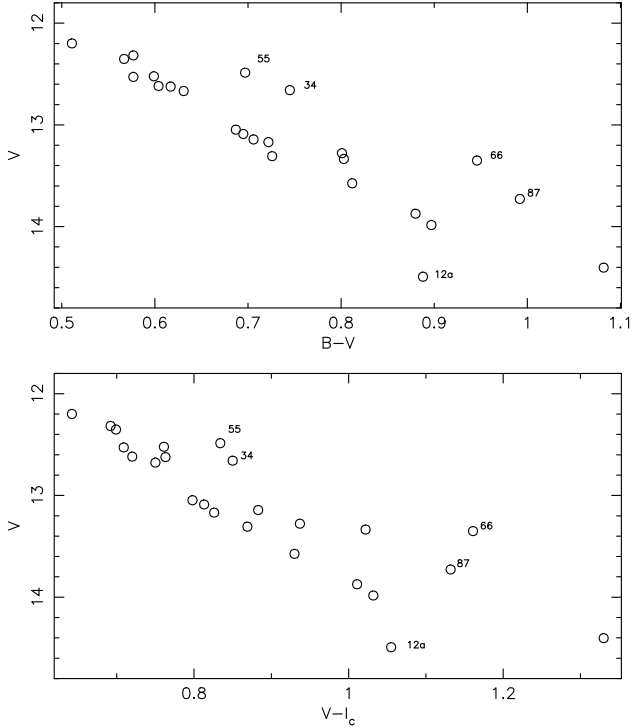}  
\caption{Colour magnitude diagrams for the sample of stars chosen for
high resolution spectroscopy in NGC 2547. A number of stars referred to in section
2.3 are labelled. (a) $V$ vs $B-V$. (b) $V$ vs $V-I_{\rm c}$.}
\end{figure}

The stars we have observed spectroscopically
are indicated by the circled points in Fig.1. The colours and
magnitudes of these stars are listed in Table~1, where
we adopt the identifiers and data from JT98. Some of these stars lie
significantly above the single star sequence in the $V,B-V$ and
$V,V-I_{\rm c}$ colour-magnitude diagrams (CMDs) 
shown in Fig.2, possibly due to binarity.
Where a target is 0.4 mag
or more brighter in $V$ than suggested by its intrinsic $B-V$ and
$V-I_{\rm c}$, we have identified it as a {\em possible} binary that is
unresolved by the CCD photometry in JT98. Note that the presence of an
unresolved binary with period greater than a few days is unlikely to
affect X-ray activity as measured by $L_{\rm x}/L_{\rm bol}$ (although
such sources should be easier to detect because of their higher $L_{\rm
x}$), but that shorter period binary systems may have enhanced X-ray
activity over single stars of the same colour because of tidally
enforced rapid rotation.

In summary,
we believe our sample is
unbiased with respect to rotation for $(V-I_{\rm c})_{0}>0.8$, but may 
be missing some of the slowest rotators in a truly optically selected sample
for stars hotter than this.

\subsection{High resolution spectroscopy}

\begin{figure}  
\vspace{13.8cm} \includegraphics{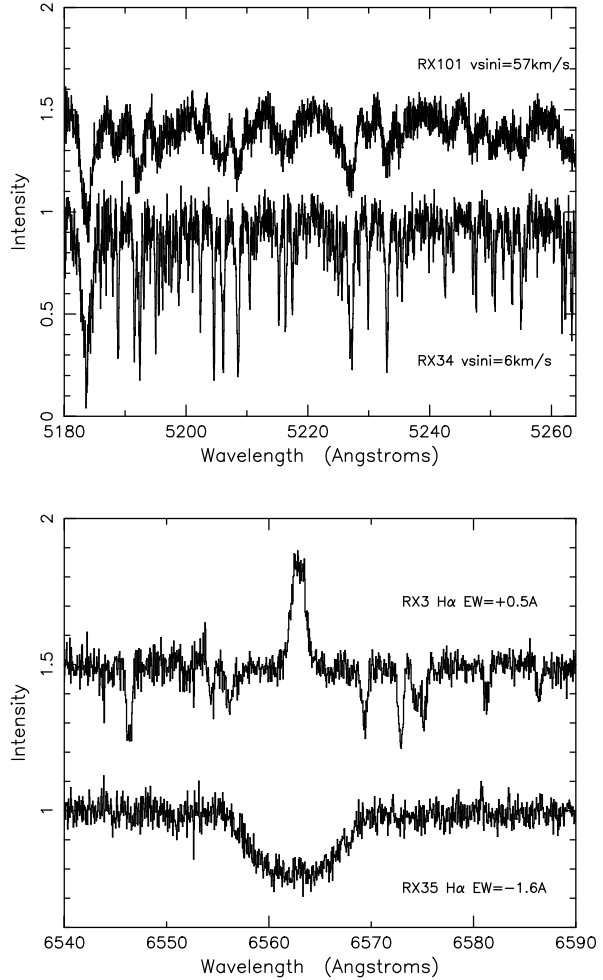}  
\caption{Examples of our high resolution spectra: (a) One of the orders
used for radial velocity and \vsini\ measurements. (b) Spectra around H$\alpha$ for
two of our targets, including the ultra-fast rotator RX35.}
\end{figure}

The stars circled in Fig.1 and shown in Fig.2 were observed on the 6th and 7th of
January 1999 at high resolution using
the UCLES coud\'{e} echelle spectrograph on the 3.9-m Anglo-Australian Telescope.
The data were collected using a 79 grooves/mm echelle grating and a
4096 by 2048 pixel MIT/LL CCD device. Each cross-dispersed echellogram
covered a wide spectral range (but with gaps between orders) from about
5000\AA\ to 8500\AA,
including the H$\alpha$ and O\,{\sc i} 6300\AA\ lines. There were several
orders with no telluric lines and many neutral metal lines that could
be used to measure radial velocities and projected equatorial velocities by
cross-correlation with standard stars. The 1.2 arcsec slit projected to
about 3.5 pixels on the CCD leading to a resolving power of around
44000 (or a resolution of 0.15\AA\ at H$\alpha$). A dekker was used to
separate the orders, but the spatial width of each order (about 14
arcsecs) was sufficient to achieve background subtraction, given the
typical 1.3-1.6 arcsec seeing that was encountered during the majority of
the run.

Targets were observed for between 20 minutes and 1 hour, resulting in
consistent signal to noise levels of about
20-25 per CCD pixel. Along with the usual flats, bias and arc lamp
exposures, we also obtained spectra of several radial velocity
standards, a number of slowly rotating
spectral type standards with minimal chromospheric activity, 
and a number of rapidly rotating B-stars to
facilitate telluric line correction around the H$\alpha$ and O\,{\sc i}
6300\AA\ lines.

Heliocentric radial velocities (RVs) were determined for all our targets by
cross-correlation with the spectra of RV standards -- HR 4786, HD
4128 and HD 126053. Two orders with spectral ranges
$\lambda\lambda5157-5282\AA$ and $\lambda\lambda5989-6141$ were used
independently with each standard.
The average value is taken from all these cross-correlations and the
standard deviation used as an estimate of the likely RV error. This is
typically about 1\kms, although it is higher for very fast rotators. The
internal consistency of the measured standard RVs suggests our external
error is of order 1\kms. The RV results are presented in Table~1.

Projected equatorial velocities (\vsini) were estimated by
cross-correlation with slowly rotating stars of similar colour
to the targets. The stars used were HD 102870 (F9V), HD 126053 (G1V)
HD 115617 (G5V), HD 10700 (G8V), HD 4628
(K2V) and HD 16160 (K3V). The FWHM of the cross-correlation function
yields \vsini\ when calibrated by broadening the slowly rotating
standards with limb-darkened rotational profiles of increasing \vsini.
At the S/N levels of our target spectra, we find that \vsini\ becomes
unresolved below 6\kms, has a $\sim1$\kms\ error up to 10\kms\ and about
a 10\% error in \vsini\ for more rapid rotators. The rotational
velocities determined in this way are listed in Table~1.
Figure 3a shows two examples of our spectra for both rapid and slowly
rotating stars.

In all but two cases the RV cross-correlation functions were strong and
single-peaked. RX55 is a probable SB2 binary system, with two
clearly resolved narrow peaks in the cross-correlation functions. A \vsini\
upper limit was determined for each component. RX35 seems to be an
example of an ultra-fast rotator. No clear peak could be seen in the RV
cross-correlation functions and the only feature of note visible in
the spectrum is a very broad H$\alpha$ absorption line (see Fig. 3b). 
Assuming the width of this
line was due to rotational broadening, we estimate that
$\vsini>160$\kms. The centroid of the H$\alpha$ line was used to determine the
approximate RV listed in Table~1. The same technique applied to the
next fastest rotator, RX30a, yields a \vsini\ of 110\kms and an RV of
$9\pm8$\kms, in reasonable agreement with the cross-correlation results.

The equivalent width (EW) of the H$\alpha$ line was measured by integration
under a continuum modelled with a third order polynomial. Telluric
features were divided out approximately by reference to a rapidly
rotating B-star spectrum. The EWs are listed in Table~1. For relatively
slow rotators ($\vsini<30$\kms), the width of the H$\alpha$ feature is
roughly constant and the EWs are accurate to about 0.05\AA. For the
broader rapid rotators, the errors are more like 0.1-0.2\AA\ and 0.3\AA\
for RX35. Figure 3b shows two examples of our H$\alpha$ spectra.

\begin{table*}  
\caption{Spectroscopy in NGC 2547. Identifiers and photometry are from
JT98. Columns 5 and 6 give heliocentric RVs and \vsini. Column 7 gives
the EW of the H$\alpha$ line and column 8 lists the $L_{\rm x}/L_{\rm
bol}$ from JT98.}   
\begin{tabular}{lcccccccl}  
ID & $V$ & $B-V$ & $V-I_{\rm c}$ & RV & \vsini & H$\alpha$ EW & 
${\rm Log}( L_{\rm x} / L_{\rm bol} )$ & Notes \\   
   &     &       &             & \multicolumn{2}{c}{(\kms)} & (\AA) && \\
RX3   &14.403&1.082&1.330&$+13.4\pm0.8$&18.8&+0.5&-3.30& \\
RX10  &13.984&0.897&1.032&$+13.3\pm0.7$&20.1&-0.2&-3.32& \\
RX12a &14.492&0.888&1.055&$+29.3\pm0.5$&35.5& 0.0&-3.42& \\
RX12a &      &     &     &$+27.1\pm1.1$&35.5& 0.0&     & 2\\
RX16  &12.199&0.511&0.642&$+12.8\pm0.6$& 7.7&-1.4&-4.47& \\
RX24  &13.307&0.726&0.869&$+12.4\pm1.0$&31.2&-0.1&-3.35& \\
RX29a &12.677&0.631&0.750&$+13.7\pm0.7$&27.2&-1.0&-3.63& \\
RX30a &12.317&0.577&0.692&$+15.4\pm3.3$&86.0&-1.3&-3.43& \\
RX34  &12.658&0.745&0.850&$+12.1\pm0.6$& 6.1&-0.9&-4.45& 1 \\
RX35  &13.278&0.801&0.937&$+21\pm10   $&$>160$&-1.6&-3.60& 3\\
RX42  &12.522&0.599&0.761&$+14.2\pm0.9$&57.3&-0.9&-3.52& \\
RX49  &13.335&0.803&1.022&$+13.0\pm1.0$&$<6$&-0.4&-3.40& \\
RX52  &12.528&0.577&0.709&$+12.8\pm0.6$&28.4&-1.2&-4.01& \\
RX53  &13.574&0.812&0.930&$+13.6\pm1.1$&11.8&-0.4&-3.50& \\
RX55$^{1}$&12.486&0.697&0.834&$+47.9\pm1.0$&$<6$&-1.1&-4.38& 1,4 \\
RX55$^{2}$&      &     &     &$-17.6\pm1.1$&$<6$&    &     &\\
RX58  &13.873&0.880&1.011&$+13.3\pm1.1$&12.0&-0.3&-3.81& \\
RX64  &12.618&0.604&0.720&$+13.9\pm0.7$&23.8&-1.2&-4.28& \\
RX66  &13.350&0.946&1.161&$+19.2\pm1.1$& 6.0&+0.1&-3.30& 1\\
RX69  &13.169&0.722&0.826&$+12.1\pm1.1$&12.8&-0.6&-3.78& \\
RX72a &12.623&0.617&0.763&$+11.2\pm1.1$&14.5&-0.9&-3.76& \\
RX79  &12.352&0.567&0.699&$+11.8\pm1.2$&38.5&-1.0&-3.97& \\
RX87  &13.728&0.992&1.132&$+9.3\pm1.0$ &14.7&+0.1&-3.39& 1 \\
RX94  &13.089&0.695&0.813&$+11.6\pm1.1$&29.3&-0.2&-3.31& \\
RX99  &13.047&0.687&0.798&$+11.5\pm1.1$&10.2&-0.8&-3.89& \\
RX101 &13.143&0.706&0.883&$+11.3\pm1.1$&56.5&-0.2&-3.42& \\
&&&&&&&&\\
\multicolumn{9}{l}{1 Possible photometric binary system.}\\
\multicolumn{9}{l}{2 Spectra were taken on both 06 and 07 January
1999 for RX12a.}\\
\multicolumn{9}{l}{3 RV and \vsini\ values taken from the H$\alpha$
line.}\\
\multicolumn{9}{l}{4 SB2 system. H$\alpha$ is not quite resolved so
only one
value is given for the system.}\\

\end{tabular}  
\end{table*}  

\subsection{Cluster membership}

Based on the size of the X-ray error circles and spatial density of
possible optical counterparts, JT98 estimated that there would only be 0.8
spurious correlations with X-ray sources for $V<15$ along the NGC 2547
CMD sequence. As X-ray emitting field stars at this level of $L_{\rm
x}/L_{\rm bol}$ should be relatively rare, then all but perhaps one of
our targets should be NGC 2547 members. The RV measurements offer us an
opportunity to check membership credentials. In a cluster, one expects
the single stars to have a single RV value (for clusters with
small angular extent) with a dispersion of no more than 1\kms\ or so.
Binary stars will introduce some scatter. Wide binaries will have RVs
close to the mean single star value, close binary members could have
RVs radically different from the cluster mean and SB2 systems should have two
measured RVs, either side of the cluster mean.

Consulting Table~1 we can see that 20/24 stars have RVs within 2.5$\sigma$ of a
weighted mean value of $+12.8$\kms, with a standard deviation of 0.9\kms.
The error in this mean cluster RV is dominated by the
$\sim1$\kms\ external error in the RV values. Of the other 4 stars, we
classify RX12a as a non-member, because we have two consistent radial
velocity measurements on two different nights that are very different
to the cluster mean. 
The only way we could be mistaken is if RX12a
were a cluster SB1 with an
orbital period close to the 1 day separation of the observations.
However, we also note that RX12a actually lies slightly {\em below} the single
star sequence in Fig.2, casting further doubt on a cluster SB1 interpretation.
Because RX12a has a \vsini\ of 35\kms\ though, it must be the counterpart
to the X-ray source.
Both RX66 and RX87 are likely to be SB1 cluster members in
moderately wide binary systems because both lie significantly above the
trend of single stars in Fig.2 and both have RVs which are only
3-6\kms\ from the cluster mean. They are also both chromospherically
active in the H$\alpha$ line (see section 3 and Table 1) and therefore
almost certainly the counterparts to the X-ray sources.
Of course we cannot rule out that they
are in fact short period systems with large velocity amplitudes or
indeed interloping field stars with very high X-ray emission -- these
possibilities are simply less likely. RX55 is obviously an SB2, with
roughly equal mass components judging by the nearly equal power in each
of the cross-correlation peaks, and by the fact it lies about 0.7 mag
above the cluster sequence in Fig.2. The mean RV of the components is
$+15.1\pm1.5$, very close to the cluster mean. We conclude that this is
a nearly equal mass SB2 member of the cluster. If the velocity
amplitude is at least 32\kms\ and the components are 1\msun\ then the
period is less than $74\sin^{3}i$ days, where $i$ is the (unknown)
orbital inclination. Without further information we cannot say whether
the period of RX55 is short enough to tidally enforce rotation at the
orbital period. The \vsini\ upper limits for the two components could
be perfectly consistent with rotation periods synchronised to
orbital periods of a few days if the system has a low orbital and rotational
inclination. Finally we note that RX34 is also a possible binary system
based on its position in Fig.2, but if so, it is probably a
moderately wide binary system because its RV agrees with the cluster mean.

In summary, we have strong evidence that 20 of the 24 stars we have
observed are single cluster members, 2 are possible SB1 members, 1
is almost certainly an SB2 member and 1 is probably a very active field
star behind the cluster.
  
\section{Results}  

\begin{figure}  
\vspace{7.5cm} \includegraphics{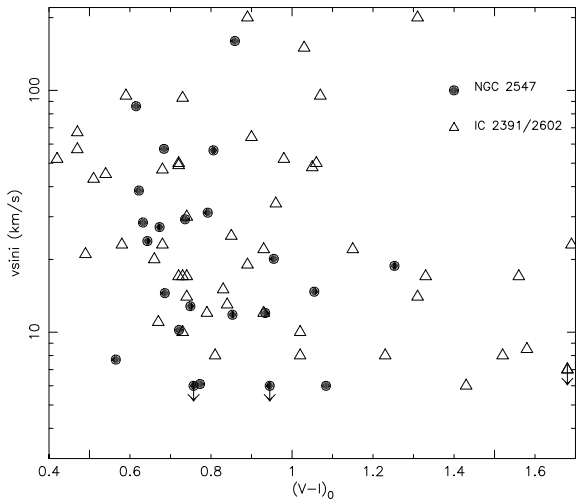}  
\caption{Projected equatorial velocities (\vsini) as a function
of intrinsic colour for NGC 2547 (dots) and IC 2391/2602
(triangles).}
\end{figure}  

\begin{figure}  
\vspace{9.0cm} \includegraphics{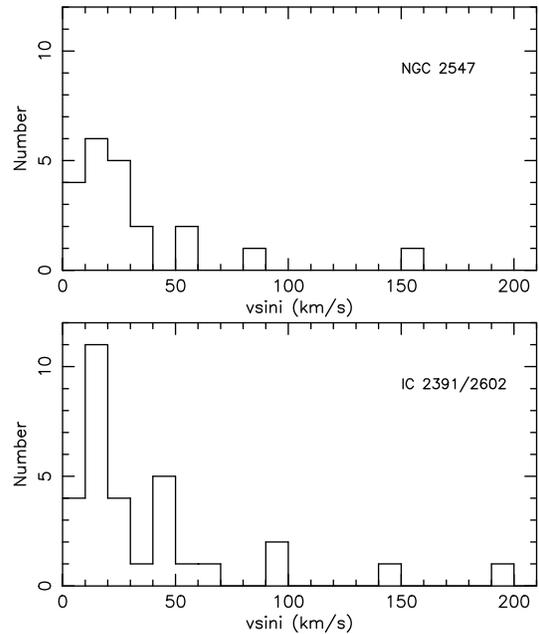}  
\caption{Histograms of \vsini\ for solar-type stars in NGC 2547 (top) and IC
2391/2602 (bottom) in the colour range $0.6<(V-I_{\rm c})_{0}<1.1$.}
\end{figure}

The best way to present our results is by comparison with the IC 2391
and IC 2602 clusters. These have been studied at X-ray wavelengths by
Randich et al. (1995), Patten \& Simon (1996) and Simon \& Patten
(1998) and have been the subject of an extensive optical
spectroscopy campaign to establish membership and measure rotational
broadening and chromospheric activity by S97. Rotation periods (as
opposed to \vsini) have also been measured in samples of cool stars
in IC 2391 and IC 2602 by Patten \& Simon (1996) and Barnes et
al. (1999) respectively.

Our first exhibit is Fig.1, which as we have said previously
shows that the X-ray activity in all the clusters rises rapidly between
$(V-I)_{0}\simeq 0.5-0.7$, but that the coronal activity peaks in IC
2391 and IC 2602 at $L_{\rm x}/L_{\rm bol}\simeq10^{-3}$, about a factor of
two higher than in NGC 2547. The extra information we now have is that
all but one of the circled NGC 2547 stars are probable cluster members,
so any reduced X-ray activity levels cannot be blamed on field star interlopers.
Exactly the same picture is seen in a
comparison with the cool stars of the 
older Alpha Per cluster (this is Fig.13 in JT98 --
which we do not reproduce here), but this and Fig.1 also clearly
show that at late spectral types (($V-I_{\rm c})_{0}>1.5,
(B-V)_{0}>1.3$), the peak level of X-ray emission in NGC 2547 returns
to a level that is indistinguishable from that in the IC 2391/2602, Alpha Per or Pleiades
(see Stauffer et al. 1994) late K and M-stars.

Figure 4 compares projected equatorial velocities in the clusters
as a function of intrinsic colour. We are more comfortable comparing
\vsini\ than attempting a comparison with the measured rotation periods
in IC 2391/2602 (Patten \& Simon 1996; Barnes et al. 1999) because (a)
we do not have rotation periods for NGC 2547 and (b) rotation
period data can be subject to additional selection biases beyond those we
discuss immediately below, because of the need for substantial magnetic spot
activity on the stellar surface and possible sampling biases.

Recall that we believe the NGC 2547
sample is unbiased with respect to rotation for $(V-I_{\rm c})_{0}>0.8$, but
may lack some of the slowest rotators among the hotter stars. Note that
in Fig.4 and subsequent figures, we have not included the non-member
RX12a and display the SB2 system, RX55, as a single point. 
S97 have used similar arguments to those that we used in section
2.1 to show that their X-ray selected sample of G-stars in IC 2391/2602 should be
unbiased with respect to rotation, but that hotter and cooler stars may
preferentially be the more rapidly rotating cluster members.
Figure 4 suggests that
the rotation rate distributions are extremely similar,
although perhaps
there are one or two more slow rotators in NGC 2547 than in IC
2391/2602 where all the solar-type stars have resolved
$\vsini\geq8$\kms.
However, given the uncertainties in inclination angles and
sample completeness at low rotation
rates we do not think there is strong evidence for differences in the
slowest rotation rates in NGC 2547 and IC2391/2602.
Figure 5 shows the \vsini\ histograms in NGC 2547 and IC 2391/2602 
for an intrinsic colour range
$0.6<(V-I_{\rm c})_{0}<1.1$ that encompasses the bulk of the NGC 2547
sample. As the reader can see, the distributions are very similar.
We have performed a formal Kolmogorov-Smirnov double-sided test between the
cumulative $\vsini$ distributions which yields a probability of only 28\%
that the NGC 2547 and IC 2391/2602 distributions are drawn from
differing parent samples.

\begin{figure}  
\vspace{7.5cm} \includegraphics{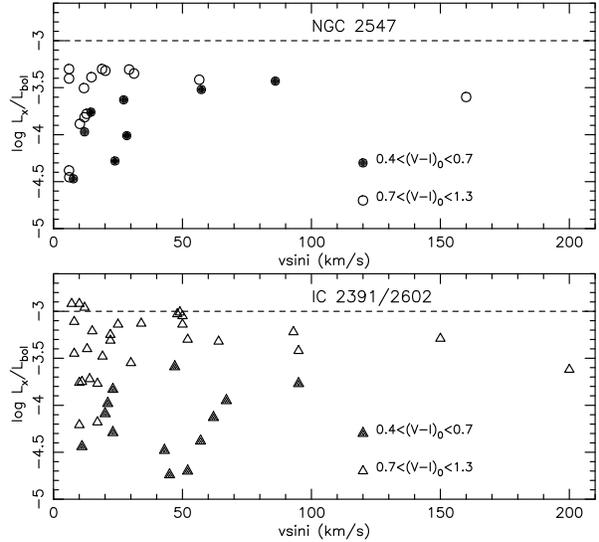}  
\caption{The X-ray activity-rotation correlations for NGC 2547
(top) and IC 2391/2602 (bottom). The samples have been split into two
intrinsic colour bins: Solid symbols are for $0.4<(V-I_{\rm c})_{0}<0.7$ and
open symbols for $0.7<(V-I_{\rm c})_{0}<1.3$.}
\end{figure}  

\begin{figure}  
\vspace{6.2cm} \includegraphics{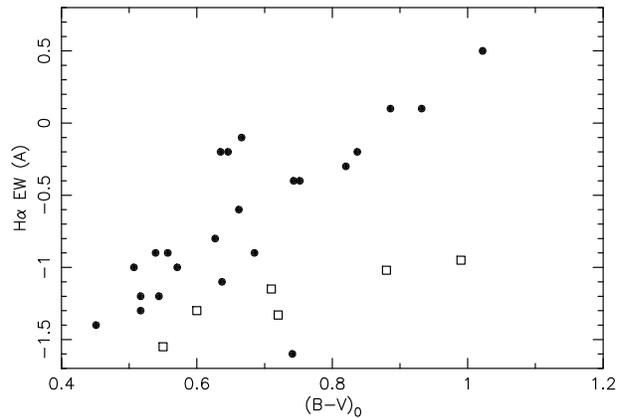}  
\caption{ H$\alpha$ equivalent width versus intrinsic $B-V$ for
NGC 2547 (dots) and a set of slowly rotating, low activity stars that
we used for \vsini\ cross-correlation templates (squares).}
\end{figure}  

\begin{figure}  
\vspace{6.2cm} \includegraphics{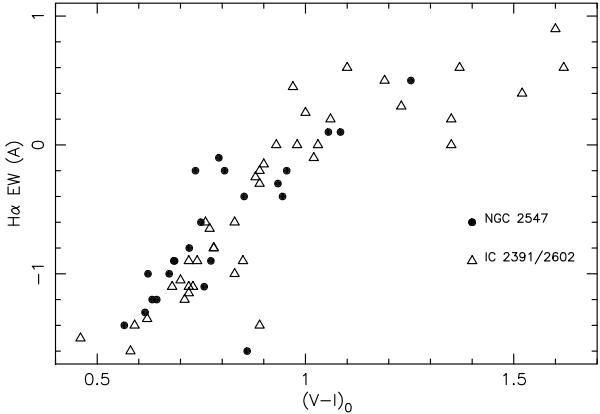}  
\caption{H$\alpha$ equivalent width versus intrinsic $V-I_{\rm c}$ for
NGC 2547 (dots) and IC 2391/2602 (triangles). The levels of
chromospheric activity appear very similar in the three clusters.}
\end{figure}

What is very clear from Figs. 4 and 5, is that NGC 2547 does not appear to
contain an anomalously slowly rotating population of solar-type stars.
There are several
late F and G stars with $v_{e}\,\sin i\geq20$\,km\,s$^{-1}$, 
the threshold for X-ray
saturation in these stars defined by Stauffer et al. (1994), and 4
ultra-fast rotators with $\vsini>50$\kms.

Figure 6 combines the rotation and X-ray data for NGC 2547 to show the
the rotation-activity correlation compared
with that in IC 2391 and IC 2602. We have split the samples according to their
colour. There is a weak correlation present for stars with
$(V-I)_{0}<0.7$ in NGC 2547 and very little sign of a correlation in IC
2391 and IC 2602. We think this is because in F stars, the depth of the
convection zone, rather than rotation rate is the dominant influence on
dynamo activity. If there were enough stars in these clusters, and the
photometry was accurate enough, we believe that choosing a very narrow
colour range {\em would} yield a rotation-activity correlation.
This is made clearer by Fig. 1, where the X-ray
activity is seen to rise steeply between $0.5<(V-I)_{0}<0.7$. Small differences in
colour and the randomizing effect of unknown rotation axis inclination angles
can effectively destroy the correlation with rotation in this colour interval.

For cooler stars $(0.7<(V-I_{\rm c})_{0}<1.3)$ it does appear
that X-ray activity rises with rotation rate and saturates above a
$v_{e}\,\sin i$ of about 
15-20\,km\,s$^{-1}$. The correlation is not perfect, probably
because of random inclination angles. Stars with low \vsini\ could have
high activity levels but be fast rotators viewed close to pole-on.
This may well be the case for RX49 and RX66.
The key point is that NGC 2547
apparently saturates at a lower coronal activity level than IC
2391 and IC 2602, but within the limitations of the few data points we
have, the saturation seems to occur at a similar rotation rate.
There is also weak evidence, based only on RX35, that as in the IC2391/2602
and Alpha Per clusters, there may be a decrease in the saturated level
of X-ray emission at very fast rotation rates. 

Figures 7 and 8 show the behaviour of H$\alpha$ as a
function of colour in the three clusters, represented by the equivalent
width (EW) of the H$\alpha$ feature. In magnetically inactive stars, H$\alpha$ is
always seen in absorption, but is chromospherically 
filled in and then goes into weak emission (EW$\simeq 1-5$\AA)
for cooler stars in young open clusters. Figure 7 makes the comparison
between the H$\alpha$ EWs in NGC 2547 and the inactive standard stars
we used for the \vsini\ comparisons. The H$\alpha$ absorption line is
clearly filled or even in emission for NGC 2547 when compared with
inactive stars of the same colour. 

The three NGC 2547 stars that lie above the trend for the other cluster
members in Fig.7 are RX24, RX94 and RX101. These objects do
indeed have the largest \vsini\ values in this colour range. However,
the correlation with rotation is certainly not perfect. RX42, with a
\vsini\ of 57\kms\ and a colour only slightly bluer than these three
objects lies in the sequence defined by the majority of stars. 
Another peculiarity is that the ultra-fast rotator RX35 appears to have 
anomalously deep H$\alpha$ absorption for its colour, deeper even than inactive,
slowly rotating stars of the same colour. We have no convincing
explanation for this strange result at present. We can speculate
though, that perhaps RX35 is seen almost equator on and is surrounded
by the cool, ``slingshot prominences'' that have been seen around some ultra
fast rotating field stars and G stars in the Alpha Per cluster
(Collier-Cameron \& Robinson 1989; Collier-Cameron \& Woods 1992;
Jeffries 1993). These prominences appear to be co-rotating clouds,
confined by the stellar magnetic field, which scatter chromospheric
H$\alpha$ photons out of the line of sight -- causing absorption
features which move from blue to red across the stellar H$\alpha$ profile.
Our single 30 minute observation (see Fig.3b) may have been too long to resolve
individual cloud features, or there may be many clouds around the star,
because the H$\alpha$ profile appears reasonably smooth. If this
explanation were true then a highly variable H$\alpha$ profile might be
expected and our \vsini\ determination based on the width of the
H$\alpha$ absorption may underestimate the true \vsini.

Figure 8 compares NGC 2547 with IC 2391/2602. 
S97 and others have suggested
that the colour at which chromospheric H$\alpha$ emission rises above
the continuum may be an excellent indicator of rotation rates and by
implication, age, because it occurs at cooler colours in the older
Pleiades and Hyades clusters.
Stars with active accretion
discs however (the CTTS), show H$\alpha$ emission way
above that seen in even the most chromospherically active stars, with
emission EWs$>10$\AA.
This diagnostic of accretion correlates excellently with
others such as near infrared excess emission over that expected from a
photosphere alone, veiling of the optical spectrum by continuum emission from
accretion hotspots or emission from forbidden metallic lines such as O\,{\sc i}
at 6300\AA\ (Hartigan et al. 1990, Hartigan, Edwards \& Ghandour 1995).
Figure 8 demonstrates that NGC 2547 behaves very
similarly to IC 2391 and IC 2602. H$\alpha$ emission first appears at
$(V-I)_{0}\simeq 1.0$, which is more or less as expected for a very
young cluster containing magnetically active stars. 
None of the stars show any signs of excess H$\alpha$ that come even
close to the levels expected from CTT accretion phenomena.

We have also checked our optical spectra for veiling or the presence of
an O\,{\sc i} emission line at 6300\AA. Edwards et al. (1993) show that infrared
excesses are present in stars with O\,{\sc i} 6300\AA\ EWs of between
0.1 and 10\AA\ and that such stars also show an excess optical
continuum ranging from 10\% to $>90\%$ of the observed flux. In our
slowly rotating stars, we can
place firm upper limits of $<0.05\AA$ on the EW of any O\,{\sc i}
emission and by comparison with the low activity standard stars we find
no evidence for any veiling continuum above a level of about 10-20\% of
the observed flux. We do not detect emission lines or veiling in the
fast rotators ($>20$\kms) either, but here the limits are relaxed to
$<0.1$\AA\ and 30\% respectively.
In summary, we can categorically state that none of
the stars we have observed show any evidence for active accretion
discs.

\section{Discussion}

\subsection{The age of NGC 2547}

How we interpret the rotation data for NGC 2547 depends a great deal on
what we assume its age is. JT98 obtained $14\pm4$\,Myr from low mass
isochrone fits. The same isochrones would yield ages of 25\,Myr for
IC2391/2602, 50\,Myr for Alpha Per and 90\,Myr for the Pleiades, in
reasonable agreement with the traditional nuclear turn-off ages
determined from high mass stars in the Hertzsprung-Russell diagram
(Mermilliod 1981). In the last couple of years these ages have been
challenged by measurements of the lithium depletion boundary (LDB) in very
low mass cluster stars. The luminosity at which lithium remains
unburned in a fully convective star that is contracting towards the
ZAMS can be mapped onto an age with reasonable precision (see
Ushomirsky et al. 1998). This method has been used to obtain ages of
$53\pm5$\,Myr for IC 2391 (Barrado y Navascues, Stauffer \& Patten
1999), $85\pm10$\,Myr for Alpha Per (Stauffer et al. 1999) and
$125\pm8$\,Myr for the Pleiades (Stauffer, Schultz \& Kirkpatrick
1998). This older age scale implies a modest amount of convective core
overshoot to bring the nuclear turn-off ages into agreement (Mazzei \&
Pigatto 1988).

Jeffries et al. (2000) have attempted to find the LDB
in NGC 2547, but could only establish a {\em lower} limit to the age
of about 23\,Myr. If the relative positions of the low mass stars in 
IC 2391 and NGC 2547 are
accepted as an indication of an age difference between the two clusters
and we assume that the LDB age for IC 2391 is correct, then NGC 2547
could be as old as 35-40\,Myr. This would make a substantial difference
in the interpretation of the rotation data because at 40\,Myr, a
solar-type star would have completed the vast majority of its
contraction (and consequent change in moment of inertia) towards the
ZAMS. At 15\,Myr, the surface rotation rate could still increase by
about a factor 1.8 (neglecting angular momentum loss) due to changes in radius
and moment of inertia (from the models of D'Antona \& Mazzitelli 1997).
Also of course, any deductions about the lifetimes of circumstellar
discs are crucially dependent upon the assumed ages of the younger
clusters.

Without  making a judgement on the relative merits of the isochronal
and LDB ages, we will need to consider the case of both the younger
and older ages scales. i.e. where the ages of NGC 2547 and IC 2391 are
roughly 15\,Myr and 25\,Myr and where they are roughly 40\,Myr and 55\,Myr
respectively.

\subsection{Rotational evolution of solar-type stars}    

NGC 2547 (along with the IC 2391/2602 clusters) occupies an important age position
between the older, well studied Alpha Per and Pleiades clusters and PMS
stars in star forming regions. Previous attempts to study
rotation in this age range (10-40\,Myr) have concentrated on dispersed
populations of X-ray selected objects in and around OB and T
associations ({\em e.g.} Bouvier et al. 1997b). These studies appear to
show a lack of the slow rotators that are needed to explain the older 
Pleiades rotation distribution where 50\% of solar-type stars have 
$\vsini<10$\kms\ (Queloz et al. 1998). The problems with these
investigations of scattered PMS populations 
are that the stellar ages rely on rather uncertain
distances and that X-ray selection might be quite severe, so
biasing against the presence of slow rotators. In our NGC 2547 study
and in the IC 2391/2602 study of S97, it seems likely that for
solar-type stars at least, this selection effect is absent or weak.

The rotation rates we have measured in NGC 2547 largely confirm the
results found in IC 2391/2602 by S97. If the clusters had ages of
40\,Myr and 55\,Myr respectively and assuming the initial angular
momentum distributions and circumstellar disc lifetimes were similar,
then we would expect to see little
difference in their rotation rate distributions. This is because solar type stars
in both clusters would have reached the ZAMS, there would be little
moment of inertia change and angular momentum losses over the course of
15\,Myr might be too small to be measured, except perhaps in the few most
rapid rotators. However, if the clusters
were aged 15\,Myr and 25\,Myr, then assuming solid-body rotation and
ignoring angular momentum losses, 
we might expect to see a rotational
spin-up of 50\% between NGC 2547 and IC 2391/2602.
The median \vsini\ among the solar-type stars ($0.6<(V-I_{\rm
c})_{0}<1.1$) in NGC 2547 is about 20\kms and very similar in IC 2391/2602. 
These figures are based on relatively small numbers but
one could view this similarity as a (very) weak argument for the older age
scale.

The upper quartile of rotation in the Pleiades solar type stars 
occurs for $\vsini>15$\kms. In IC 2391/2602 it is 50\kms\ and about
40\kms\ in NGC 2547. The numbers here really are too small to analyse
any difference between NGC 2547 and IC 2391/2602. If the solar type
stars in all three clusters have completed their PMS contraction,
then taken together, the results indicate that if they rotate as solid
bodies, the fastest rotating
stars must lose 60-70\% of their angular momentum between 40-50\,Myr
and $\sim125$\,Myr. If the clusters are
younger then the amount of angular momentum loss must be even greater (80-90\%)
to allow for some contraction and spin up onto the ZAMS. The similarity
in $\vsini$ of the rapid rotators of NGC 2547 and IC 2391/2602
argues for similar initial conditions and circumstellar disc lifetimes.

Overall what we have measured is in very good agreement with the models
put forward by Bouvier et al. (1997a). These
models start with an observed rotation rate in T-Tauri stars and evolve
this using wind angular momentum loss (which saturates at fast rotation
rates) and early coupling to a circumstellar disc which is responsible
for the predominance of slow rotators at later times. These solid-body
rotation models
predict maximum rotational velocities of order 150-200\kms\ for ages
15-40\,Myr and that roughly half the solar-type stars in NGC 2547 
should have $\vsini<20$\kms, which is what we have measured. 
This high proportion of slow rotators is
achieved by assuming circumstellar disc lifetimes as old as 40\,Myr in
some stars and that about 15\% of stars (the slowly rotating ones) 
are locked to their discs at 15\,Myr.

We have no evidence that any of the solar-type stars in NGC 2547 still
have discs. This could argue that the cluster has an age of
$\sim40$\,Myr, but other interpretations are possible. 
Cameron \& Campbell (1993) and Armitage \& Clarke (1996) show that
discs with mass accretion rates of only $10^{-10}$\msun/yr can still
enforce rotational equilibrium, whereas mass accretion rates at least
an order of magnitude higher are needed to provoke the optical
accretion signatures we have searched for in this paper. Some of our
slowest rotators may still have remnant discs and it would be worth
searching for these in more detail at infrared wavelengths. Another
possibility is that the radiative core and convective envelope are not
perfectly coupled and that interior differential rotation is
possible. This would allow less angular momentum loss to produce slow
rotators and thus requires shorter disc lifetimes. However, the
core-envelope coupling timescale must be substantially greater than the
10\,Myr proposed by Keppens et al. (1995), who found that short disc
lifetimes ($\sim6$\,Myr) could not produce enough slow rotators on the
ZAMS in
these circumstances. The differential rotation models put forward by
Krishnamurthi et al. (1997) show that sufficient slow rotators {\em can} be
produced with disc lifetimes of 3-10\,Myr if the core-envelope 
coupling timescale is of order 100\,Myr. 

Alternatively, NGC 2547 and
IC 2391/2602 may have been born with a population of slower rotators
than are typically seen in even younger clusters and star forming regions.
Barnes et al. (1999) show that
to explain the slow rotators in IC 2602 requires that 
20-30\% of solar-type stars in IC 2602 needed to have initial periods
as slow as 16 days if disc lifetimes are to be limited to $\leq3$\,Myr.
However, Choi \& Herbst (1996) and Kearns \& Herbst (1998) find that
90\% of stars in the 3\,Myr old NGC 2264 cluster and 1\,Myr Orion
Nebula and Trapezium clusters have rotation periods faster than 10
days. Thus either long disc lifetimes ($>10$\,Myr), 
internal differential rotation with long 
core-envelope coupling timescales ($\sim100$\,Myr), or anomalously small initial
angular momenta are required to explain the slow rotators
in NGC 2547, IC 2391 and IC 2602.

In any of these scenarios it is at least clear that the NGC 2547 (and
IC 2391/2602) data have partially solved the problem of the lack of slow rotators
at ages between the PMS stars in Taurus and Orion and the older Alpha Per
and Pleiades clusters. The slow rotators are there, it just requires
observations of optically selected samples, or at least complete X-ray
selected samples which are not biased against slow rotators. We can
also say that the combination of initial angular momentum distribution
and disc lifetimes must be reasonably similar in NGC 2547 and IC
2391/2602 in order to
produce similar \vsini\ distributions at their current ages. This is an
important result because in the disc-regulated angular momentum loss
model, even a small variation in disc locking timescales would produce
big changes in the \vsini\ distribution as solar-type stars reached the
ZAMS.

\subsection{Anomalous X-ray emission in NGC 2547}

Our main reason for performing high resolution spectroscopy in NGC 2547
was to explain the anomalously low level of X-ray emission in the most
active solar type stars in the cluster. Our working hypothesis was that
all of the NGC 2547 late F and G stars were rotating slower than
20\kms\ and hence none of them showed the saturated level of X-ray
emission seen in other young clusters. Our data conclusively reject
this hypothesis. There are several examples of very rapid rotators in
NGC 2547 and the rotation rate distribution is indistinguishable from
the IC 2391 and IC 2602 clusters.

We had also suggested that slow rotation in NGC 2547 might be caused by
long lived circumstellar discs or an unusually small amount of initial
angular momentum. Neither of these is now required and we have
evidence that none of the solar type stars in NGC 2547 possess active accretion discs.
This latter discovery in itself places an upper limit on the lifetime
of such discs (that could be detected with H$\alpha$, O\,{\sc i}
emission or optical veiling) of 40\,Myr and possibly as low as 15\,Myr
depending on what is finally concluded about the age of NGC 2547.

NGC 2547 appears to follow the ARAP discussed in the introduction up
to a point. The levels of activity are appropriate for its age and
measured rotation rates, except for the case of coronal activity in the
fastest rotators. There, the ``saturation level'' for X-ray emission is
a factor of two lower than seen in all other young clusters, but appears
to occur at similar rotation rates of 15-20\kms.

The fact that
the X-ray saturation occurs at a similar rotation rate and that the
chromospheric activity in NGC 2547 concurs with that in IC 2391 and IC 2602, leads
us to suspect that perhaps the X-ray fluxes in NGC 2547 have been
underestimated by a factor of two. A further piece of evidence in
support of this view is that RX12a, which is almost certainly a
background field star, has $\vsini=35.5$\kms and $L_{\rm x}/L_{\rm bol}=10^{-3.42}$.
The spectral type of this object (determined from the
cross-correlations) is about K0V, so it does not appear to suffer from
much more reddening or absorption than the NGC 2547 stars. If the low
X-ray saturation levels in NGC 2547 are peculiar to that cluster for some
reason, then it is difficult to explain why a similar phenomenon 
occurs in an unconnected fast rotating field star in the
same direction.

A factor of two underestimate of intrinsic X-ray fluxes might be possible if the
interstellar absorption towards the cluster is higher than assumed by
JT98, or if the assumed stellar X-ray spectrum is incorrect. Another
factor to consider is that the IC 2391, IC 2602, Alpha Per and Pleiades
{\em ROSAT} data we have discussed were obtained with the Position
Sensitive Proportional Counter (PSPC), rather than with the High Resolution
Imager (HRI) as in the case of NGC 2547. However, we do not believe there
could be as much as a factor of two error produced here. The assumed 
interstellar column density for NGC 2547 of $3\times10^{20}$\,cm$^{-2}$ was
estimated from the cluster reddening (Bohlin, Savage \& Drake 1978) 
and from Lyman $\alpha$ measurements of
early type stars at similar galactic coordinates (Fruscione et
al. 1994). A likely error in these estimates is 50\%, but the column
density would have to be as high as (2-3)$\times10^{21}$\,cm$^{-2}$
to double the unabsorbed X-ray fluxes in NGC 2547.
Such a large column density probably exceeds the 
column density out of the Galaxy in this
direction, as deduced from 21\,cm maps (Marshall \& Clark 1984).
JT98 also assumed a 1\,keV optically thin plasma, but again, the
consequences of this coronal temperature being wrong by as much as 50\%
only changes X-ray fluxes by 15\% (David et al. 1996). We also cannot
appeal to a mismatch in the calibrations of the PSPC and HRI.
Simon \& Patten (1998) have measured X-ray fluxes for stars in IC 2391
with the HRI and compared them with fluxes measured for the same stars
with the PSPC by Patten \& Simon (1996). They find excellent agreement
with essentially no systematic difference and little variability in the
X-ray fluxes. The HRI count rate to flux conversion factor used by
Simon \& Patten (1998) is 10\% smaller than that used for NGC 2547 by
JT98, but this is consistent with the smaller assumed value of
interstellar absorption for IC 2391. Lastly, we can also say that the
bolometric corrections used by various authors to calculate $L_{\rm
x}/L_{\rm bol}$ are the same to within a few hundredths of a magnitude
at all colours, so 
that none of the discrepancy can be attributed to differences in these.

Our other line of argument for claiming that JT98 calculated the fluxes
correctly is more indirect. It seems that the peak levels of X-ray
activity in the M stars of NGC 2547 are very similar to those in IC
2391/2602 and other clusters. Fig.1 shows that this is the case. Note
that we have not tried
to compare mean X-ray emission in these cool stars
because the samples are incomplete and heavily
biased towards the most active stars. A global factor of two
increase in all the X-ray fluxes in NGC 2547 would shift its M dwarfs
to activity levels higher than seen in other clusters. The only way of
escaping this problem would be if the solar type stars were more
absorbed or had radically different coronal spectra to the M
dwarfs. Neither of these seem likely but we will be able to rule them
out when we have X-ray spectra from the Chandra or XMM satellites.

We have also considered compositional
differences between the clusters as a possible solution.
Different metallicities could affect convection zone depths and
dynamo activity or may simply alter the coronal abundances and emission
measure distributions. The metallicities of these three young open
clusters are not expected to strongly depart from solar values but they
remain undetermined at the present time.
However, we have interpreted the strong turn-on of X-ray activity at
$0.5<(V-I)_{0}<0.7$ as due to the onset and deepening of convection
zones in F stars. The fact that this occurs at a very similar intrinsic
colours in NGC 2547, IC 2391 and IC 2602 argues that any compositional 
differences between the clusters are small and do not greatly influence
the magnetic dynamo efficiency.

We are left with a puzzle, perhaps akin to the very different
X-ray luminosity functions in the older (600\,Myr) 
Hyades and Praesepe clusters (Randich \&
Schmitt 1995; Barrado y Navascues, Stauffer \& Randich 1998), 
which also remains unexplained and does not seem likely
to result from very different rotational properties. That clusters with
similar ages can have different X-ray properties means we must be
careful about using X-ray data and the ARAP to draw conclusions about the age
distributions of arbitrary samples of stars, whether in clusters or the
field.

\section{Summary}  
  
We have obtained high resolution spectroscopy for a sample of
solar-type stars in the young open cluster, NGC 2547. We have
determined projected equatorial velocities, searched for the presence
of active accretion discs and measured their chromospheric activity.
Our main conclusions can be summarised as follows.
  
\begin{itemize}  
  
\item The rotation rate distribution in NGC 2547 is indistinguishable 
from that in the slightly older IC 2391 and IC 2602 clusters. In the
current paradigm for the rotational evolution of cool stars, this
points to very similar initial conditions and circumstellar disc
lifetimes in the three clusters.
  
\item We find both examples of ultra-fast rotating stars and stars with
$\vsini<10$\kms. If the slow rotating stars evolve from populations
with the rotation rate distributions seen in very young PMS clusters,
then either very long ($\sim 10-40$\,Myr) disc lifetimes or internal
differential rotation are required. An alternative might be that NGC
2547 (and IC 2391/2602) were born with a high proportion of anomalously
slowly rotating objects.

\item We find no evidence for active accretion discs in our
sample. This sets an upper limit to the lifetime of such discs at
15-40\,Myr, depending on what is assumed for the age of NGC 2547.
  
\item The slowly rotating objects in NGC 2547 (and IC 2391/2602) have
no counterparts in the X-ray selected samples of 10-40\,Myr PMS stars
that have been studied previously. We ascribe this to biases towards
fast rotators caused by strong X-ray selection. We believe this bias is
weak or absent in our sample of solar-type stars.

\item NGC 2547 appear to follow the same rotation-activity correlation 
seen in other young clusters. X-ray activity increases up to a saturated
peak for $\vsini>15-20$\kms. However,
we are unable to explain why the X-ray activity of solar-type
stars in NGC 2547 saturates at $\log (L_{\rm x}/L_{\rm bol}) = -3.3$, 
a factor of two lower than in
other young clusters. We rule out slow rotation, and consider
significant uncertainties in
calculating the X-ray fluxes unlikely.
\end{itemize}

\section*{Acknowledgements}  
  
This work was based on data collected at the Anglo-Australian
Observatory, partly funded by the UK Particle Physics and Astronomy
Research Council (PPARC). We thank the staff of the Anglo-Australian Observatory 
for assistance during the observations. Computational work was performed  
on the Keele and St Andrews nodes of the PPARC funded Starlink network.

\nocite{patten96} 
\nocite{randich96alphaper} 
\nocite{jeffries98n2547} 
\nocite{stern92} 
\nocite{randich95praesepe} 
\nocite{barrado98} 
\nocite{randich98supersat} 
\nocite{david96} 
\nocite{bohlin78} 
\nocite{randich97xrayrev} 
\nocite{jeffries99xrayrev} 
\nocite{stauffer94} 
\nocite{skumanich72}
\nocite{barnes96}
\nocite{krishnamurthi97}
\nocite{bouvier97}
\nocite{konigl91}
\nocite{shu94}
\nocite{armitage96}
\nocite{keppens95}
\nocite{cameron95}
\nocite{choi96}
\nocite{bouvier93a}
\nocite{edwards93}
\nocite{stassun99}
\nocite{stauffer97ic23912602}
\nocite{krishnamurthi98}
\nocite{strom89}
\nocite{skrutskie90}
\nocite{kenyon95}
\nocite{claria82}
\nocite{dantona97}
\nocite{randich95ic2602}
\nocite{patten96}
\nocite{simon98}
\nocite{hartigan90}
\nocite{hartigan95}
\nocite{mermilliod81}
\nocite{stauffer99}
\nocite{stauffer98}
\nocite{barrado99}
\nocite{mazzei88}
\nocite{jeffriesldb2000}
\nocite{bouvier97coy4}
\nocite{queloz98}
\nocite{cameron93}
\nocite{marshall84}
\nocite{fruscione94}
\nocite{barnes99}
\nocite{kearns98}
\nocite{collier89a}
\nocite{colliercameron92}
\nocite{jeffries93}
\nocite{ushomirsky98}

\bibliographystyle{mn}  
\bibliography{iau_journals,master}  
  
\label{lastpage}  
\end{document}